\documentclass[runningheads,a4paper]{llncs}

\usepackage{epsfig,url}
\usepackage{times}
\usepackage{amssymb}
\usepackage{amsmath}
\usepackage{amsfonts}
\usepackage{graphicx}
\usepackage{graphics}
\usepackage{color}
\usepackage{listing}
\usepackage{float}
\usepackage[ruled]{algorithm}
\usepackage{algpseudocode}
\usepackage{amssymb}
\usepackage{epsfig}
\usepackage{color}
\usepackage{xspace}
\usepackage{subcaption}
\usepackage{multirow}
\usepackage[scaled=0.85]{couriers}
\usepackage[pdfpagelabels,
           plainpages=false,
           bookmarks=true,
           hypertexnames=false,
           pdffitwindow=false,
           colorlinks=true,
           linkcolor=blue,
           citecolor=blue,
           filecolor=black,
           urlcolor=black,
           pdfauthor={},
           pdftitle={CrowdSurf: Empowering Informed Choices in the Web},
           pdfsubject={2014},
           pdfkeywords={YYY},
           pdfproducer={Latex with hyperref},
           pdfcreator={pdflatex}
          ]{hyperref}

\long\def\comment#1{}

\definecolor{red}{rgb}{1,0,0}

\newcommand{\X}{{{C\textsc{rowd}S\textsc{urf}}}\xspace}
\newcommand{\collector}{{collector}\xspace}
\newcommand{\controller}{{data analyzer}\xspace}

\newcommand{\corpcontroller}{{corporate controller}\xspace}
\newcommand{\commcontroller}{{advising community}\xspace}
\newcommand{\thirdcontroller}{{third-party advisor}\xspace}

\urldef{\mailsa}\path|{metwalley,traverso,mellia}@tlc.polito.it|
\urldef{\mailsb}\path|{stanislav_miskovic,mario_baldi}@symantec.com|

\title{CrowdSurf: Empowering Informed Choices in the Web\thanks{This work was conducted under the Narus Fellow Research Program.}}	

\begin{document}

\author{Hassan Metwalley\inst{1} \and Stefano Traverso\inst{1} \and Marco Mellia\inst{1}
\and \\ Stanislav Miskovic\inst{2} \and Mario Baldi\inst{1}\inst{2}}

\authorrunning{Hassan Metwalley et al.}

\tocauthor{Hassan Metwalley, Stefano Traverso, Marco Mellia, Stanislav Miskovic, Mario Baldi}

\institute{Politecnico di Torino, Italy\\
\mailsa\\
\and
Symantec Corp., USA\\
\mailsb\\
}

\maketitle
\begin{abstract}

When surfing the Internet, individuals leak personal and corporate information to third parties whose (legitimate or not) businesses revolve around the value of collected data.
The implications are serious, from a person unwillingly exposing private information to an unknown third party, to a company unable to manage the flow of its information to the outside world.
The point is that individuals and companies are more and more kept out of the loop when it comes to 
control private data.



With the goal of empowering informed choices in information leakage through the Internet,
we propose \X, a system for comprehensive and collaborative auditing of data that flows to Internet services.
Similarly to open-source efforts, we enable users to contribute in building awareness and control over privacy and communication vulnerabilities. \X provides the core infrastructure and algorithms to let individuals and enterprises regain control on the information exposed on the web.

We advocate \X as a data processing layer positioned right below HTTP in the host protocol stack. This enables the inspection of clear-text data even when HTTPS is deployed and the application of processing rules that are customizable to fit any need.
Preliminary results obtained executing a prototype implementation on ISP traffic traces demonstrate the feasibility of \X.

\end{abstract}

\section{Introduction}
\label{sec:intro}

Users increasingly rely on Internet services, looking for news and products, accessing social networks, organizing their life, etc. 
There are companies that base their business on the collection of personal information implicitly or explicitly embedded in the above users' activities. 
%
%
%
This results in leakage of information that users and companies prefer to keep private, in people being exposed to dubious third-party services, as well as in web companies (sometimes illegitimately) tracking their users. This phenomenon is ubiquitous, with even the major players taking part in it~\cite{facebooksued,GoogleSafari,kramer:pnas}.

Hence, users' concerns about privacy and information leakage largely increased, motivated also by recently exposed government surveillance programs. However, no means exist to control which data is handed to the web.

To this end, a common misconception is that encryption would solve the problem. Accordingly, HTTPS usage increased by 100\% each year, reaching about 42\% of web flows in June 2014~\cite{finamore:conext14}. In reality, the effects are quite different and rather exacerbate the problem. Firstly, encryption increases the value of data. Specifically, web services that deploy encryption establish a monopoly on information by precluding any other parties from deploying it, thus gaining a huge advantage in today's Internet where many businesses revolve around user information. 
Secondly, when HTTPS is deployed, users have no chance to rely on third parties to check and possibly choose which (personal) information they are sharing.

In this scenario, we advocate the need of a communication model where users are explicitly offered the freedom to i) understand  which services get their data, and ii) govern which information they are asked to exchange.
We envision a holistic and flexible solution to verify and control the information which is exchanged on the Internet, when using a web browser or running a smartphone app, whether connected to the corporate network or to a public WiFi hotspot. 

\X, presented in this paper, provides a framework for such a solution. It is designed as an open service to which anyone can easily contribute. A part of \X resides on client devices to both provide visibility on traffic and possibly act upon it (e.g., by modifying or blocking information). We conceived \X as a new layer that sits right between the application and the protocol stack, where information has not yet been encrypted. \X targets web surfing, and thus HTTP, the new ``narrow waist of the Internet''~\cite{Popa:hotnets2010}. This empowers the protection of the user's data and optional contributions to the system by users themselves.
Anyone contributes according to a personal level of expertise or convenience, from teams of security researchers who can collaboratively measure intricate signs of behind-the-scenes communications between the service providers, to novice users who can simply offer anonymized samples of their traffic or vote on the legitimacy of data leaving their devices.

Another part of \X resides on open cloud servers performing intensive processing tasks over massive datasets obtained through the contribution of volunteers. Specifically, the cloud component runs algorithms that clean and rank users' voting, index voluntarily submitted traffic, and attempts to discover unknown types of information leakage.
All of the data gathered and processed enables the cloud \X component to compute pieces of {\it advice} about the trustfulness of web services. This advice is shared with all the resident components that can leverage it to support users in taking informed decisions.
Users can create {\it rules} based on the received advice to enable fine grained control on the information flow. For instance, users can choose to block undesired services, or filter private information, or explicitly embrace third-party services.


The technology offered by \X is essential not only for individual users, but also for companies with the need to control the information entering and leaving the corporate network. Currently, companies are forced to trust the devices connected to the network and have hard time verifying the information they exchange. The so called ``BYOD'' (Bring Your Own Device) phenomenon and the reduced efficacy of traditional approaches based on firewalls and IDS's (mined by encryption~\cite{finamore:conext14}) further exacerbate the problem. In the corporate scenario, the open cloud service is replaced by a private component that, through the resident components, can impose filters to any device connected to the corporate network.

At last, \X allows third-parties to offer novel services, possibly complementing current client-server-based ones. For instance, \X could be used to enable a user to voluntarily use an accelerating proxy offered by an ISP only for specific types of traffic (e.g., when watching videos, but not when accessing her bank account). \X would dynamically forward traffic to the proxy or directly to the final destination depending on a set of rules provided by the user or by a third party and relying on advice obtained by the system.
\X could even be instrumental in enabling users to monetize on their personal information, should they decide for it, as proposed in~\cite{RiedererHotnets2011}.

\section{\X Description}
\label{sec:system}

We envision a crowd-sourced system in which users can voluntarily opt-to collaborate by providing explicit (e.g., their opinion) and implicit (e.g., traffic samples) information on the web services they use. In return they obtain information about web services. 
A \textit{\collector} -- running in the cloud -- collects information provided by the users, and it feeds an automatic {\it \controller}, which runs data mining algorithms to produce \textit{advices}.  An advice contains indications about trustfulness of web services. For instance,  the \controller can flag services collecting/leaking users' personal information, or services that children should not access, or services known to host malicious software.
A federated group of experts, the \textit{\commcontroller}, inspects the results provided by the \controller and interacts with it to generate the advices.
Following a collaborative approach similar to Wikipedia and the Electronic {F}rontier {F}oundation\footnote{\url{https://www.eff.org/}} (EFF), users are invited to increase the system's ``wisdom''.
They can be active in controlling the personal information they expose to services, or in forming the advices. And then they can voluntarily donate portions of their browsing activity, i.e., anonymized HTTP-level traces.
The advising community is supported by data mining algorithms that automatically raise flags.

The Internet offers some tools to help users to avoid disclosing personal information when browsing the web, e.g., popular browser plugins such as DoNotTrackMe\footnote{\url{http://www.abine.com/donottrackme.html}}, EFF's Privacy Badger\footnote{\url{https://www.eff.org/privacybadger}}, WoT\footnote{\url{https://www.mywot.com/}} and Ghostery\footnote{\url{https://www.ghostery.com/en/}}.
For mobile terminals, some proposals offer similar ideas~\cite{Agarwal:2013:PDM:2462456.2464460,Enck:2010:TIT:1924943.1924971}. Each targets some specific aspects of privacy leakage only.
Some leverage the idea of a crowd-sourced approach to inform users about website trustfulness.
More holistic technologies as TOR~\cite{dingledine2004tor} offer protection to users identity and from traffic inspection attacks, but they do not curb the personal information which is exposed to servers at application layer.
Similarly, companies are offering solutions to control web browsing~\cite{safebrowsing}. Yet, what they offer is unknown, and mostly un-verifiable.
\X is all of them, and none of them. \X's challenge is in offering a unified system to overcome limitation of current systems, which often do not cooperate and are dominated by manual decisions. We propose a flexible system that, based on the knowledge of the crowd and supported by automated algorithms, empowers users and companies, offering them the chance to retake the control on private data.

\vspace*{-0.2cm}
\subsection{\X System Design}
\label{sec:design}

For the design of \X we follow a short list of simple design requirements: 1) \textbf{Crowd-sourced}: we want the system to engage users to improve its effectiveness. 2) \textbf{Anonyminity}: we want \X to never offend users' privacy. Hence any contribution must be purged from any piece of personal information. 3) \textbf{Automated}: the system have to automatically process users' contributions to generate the advices. 4) \textbf{Client centric}: \X must be available on any device, as the default tool to support users' choices. 5) \textbf{Easy to use}: we want \X to be as simple and automatic as possible to allow anyone to use it.

\begin{figure}[t!]
  \centering
  \includegraphics[trim=0.1cm 3cm 0.1cm 3cm,width=0.6\columnwidth]{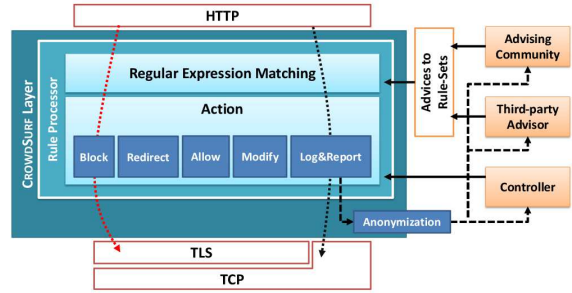}
  \caption{\X layer in the Internet stack and its high level structure.}
  \label{fig:stack}
  \vspace*{-0.4cm}
\end{figure}

Given these principles, we imagine \X cornerstone as a new layer to add to the Internet stack. We expect users' terminals, mobiles and personal computers alike, to embed the \X layer in their operating system. Fig.~\ref{fig:stack} represents the high level architecture of the \X layer. It sits between the HTTP and the transport layer, where it handles HTTP traffic, before it is eventually encrypted. This choice is motivated by the fact the today HTTP is ``the'' application layer~\cite{Popa:hotnets2010}.

Users asynchronously obtain advices from the \commcontroller, and they are free to decide to what level to take them into account: users are free to accept or overrule the notification of a potential danger. The system implements this feature through the {\tt Advices to Rule-Sets} block in Fig.~\ref{fig:stack}. It enables controlling how advices are translated in a set of \textit{rules}, or \textit{rule-set}. A rule consists of a \textit{regular expression} and one or more \textit{actions}. For each HTTP request,
the {\tt Rule Processor} looks for matches and applies the corresponding action, for instance {\tt Block}, {\tt Redirect}, {\tt Modify}, {\tt Log\&Report}, etc., with {\tt Allow} being the default one. 
This simple pattern matching/action process has proved very flexible and very efficient. It is at the root of successful technologies such as the one used in firewalls, antiviruses, traffic classifiers, etc. 

Given that \X aims at supporting a crowd-sourced approach, the {\tt Log\&Report} block is vital. It enables the collection of data samples before traffic is possibly encrypted. The layer can perform measurements at user's will and under user's control. The layer temporarily stores the measurement data locally until a certain amount is reached, at which point the layer transmits the data to the \collector. Since protecting user's privacy is strategic for \X, we adopt different approaches to avoid compromising it. The anonymization block in Fig.~\ref{fig:stack} is responsible for this.
First, it implements sampling policies, e.g., by logging only a fraction of traffic at random. This also reduces the amount of data to transfer.
Second, it filters out any piece of personal information. E.g., by default, all key values are replaced by random strings by using cryptographic hash functions. Then, a pattern/action mechanism is used. As before, the community can supply pre-defined lists of anonymization practices, which can always be customized by the user. For instance a generic policy ``remove all possible password fields'' can be augmented with ``never collect data when browsing my online bank account''.
Third, each user is assigned a unique random identifier, rotated periodically (e.g., every day).
Fourth, data on the \collector will be stored for only the time needed to process it (also to limit the storage at the \collector). 
At last, since even information available at the network layer (e.g., IP addresses) could be exploited to trace back the identity of the user, transmission to the \collector takes place through an encrypted channel established over a randomly chosen CrowdProxy, i.e., by employing other devices running the \X layer as relays.
The \collector automatically provides the identity of other \X devices among which to randomly choose the relay.

\begin{figure}[t!]
  \centering
  \includegraphics[trim=0.1cm 1cm 0.1cm 1cm,width=0.65\columnwidth]{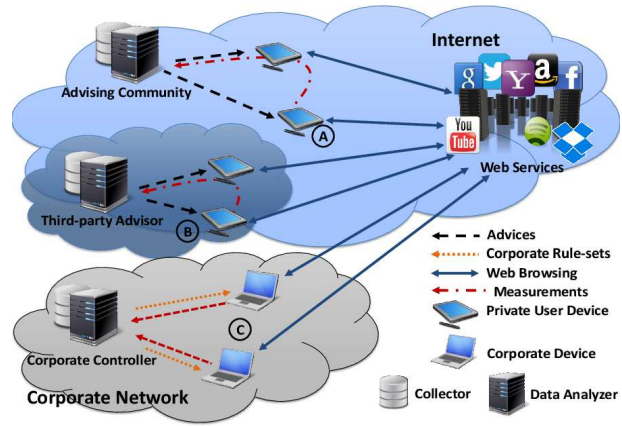}
  \caption{The entities participating to \X and their deployment for different network scenarios.}
  \label{fig:scenarios}
  \vspace*{-0.4cm}
\end{figure}

Fig.~\ref{fig:scenarios} represents possible deployment scenarios. Private user A accessing the Internet receives advices from the \commcontroller (dashed black unidirectional arrows) and possibly use them to regulate its access to web services (solid blue double-headed arrows). If A's preferences allow it, traffic samples are sent to the \collector via a CrowdProxy (dashed red arrows).

The \commcontroller and public \controller may be supported by public bodies or non-profit organization like EFF.
However, advices could also be generated by a \thirdcontroller run by an independent, third-party entity which offers custom advices to users. This  opens a ``market of advices''. For instance, user B opts for a service offered by a third party advisor. 

Finally, as shown in the bottom half of Fig.~\ref{fig:scenarios}, \X can also be deployed in a corporate scenario. In this case the \corpcontroller does not create advices, but it directly imposes rule-sets (orange dotted arrows) which are installed on devices connected to the corporate network (employee C).
Indeed, we expect the employee not to be allowed to modify the rule-sets imposed by the corporate authority.
Notice also that devices may be asked to report employees' browsing activity on administrator's demand directly to the \corpcontroller, without involving other devices. 
The presence of the \corpcontroller must be automatically identified by any device connected to the corporate network including those BYOD. This can be achieved for instance using DHCP extensions, or using standard DNS names that forces the \X layer to connect to the \corpcontroller. Notice that the same rule can be imposed on any corporate-owned device even when connected from other networks.

\begin{table}[t!]
    \centering
    \resizebox{0.45\columnwidth}{!}{
    \begin{centering}
    \begin{tabular}{c r c|c|c}
               & & \multicolumn{3}{c}{\textbf{Actions}} \\
               & & \texttt{block}    &  \texttt{redirect} & \texttt{log\&report} \\ \cline{3-5} \cline{3-5}
     \multirow{8}{*}{\rotatebox{90}{\textbf{Web Services}}}          & Facebook         & C           &              &           \\
               & Twitter          &          &                   & C           \\
               & Dropbox          &          &                   & C           \\
               & Google           &          & C($\to$Bing)      &             \\
               & YouTube          & C        &                   &             \\
               & Ebay+Amazon      & C        &                   &             \\
               & Adult Sites      & C, K      &                   &             \\
               & Trackers         & P        &                   & K           \\ 
               & Ads+NoJS         & P        &                   &           \\ 
    \end{tabular}
    \end{centering}}
    \vspace{0.3cm}
    \caption{Rules for the (P)aranoid, (K)id and (C)orporate profiles.}
    \label{tab:profiles}
    \vspace*{-0.6cm}
\end{table}

\vspace*{-0.2cm}
\subsection{\X Application Examples}
\label{sec:examples}

In the following we describe examples of \X applications in both the public Internet, and the corporate network. We use the same examples to run the experiments presented in Sec.~\ref{sec:overhead}.

A summary of rules is available in Tab.~\ref{tab:profiles}.
We define a ``Paranoid Profile'' that opts for blocking all advertisement sites, to not run Javascript code, and to use private navigation mode on the browser. This profile is the equivalent of running AdBlockPlus and NoScript plugins. This user decides to not share any traffic samples with the community.

A second profile is called ``Kid Profile'': the user activates parental control by installing the advices provided by the \commcontroller. In the experiment, we simply use the list of the Alexa top 50 ``Adult Sites'' augmented by other manually verified adult sites.
The user contributes also to manually signal other offending websites/objects he gets into. Finally, he volunteers to enable logging and reporting of the three most popular online trackers (\textit{doubleclick.net}, \textit{scorecardresearch.com}, and \textit{yieldmanager.com}).

A third profile impersonate the ``Corporate Profile'': rules are imposed by the network administrator, and i) do not allow employees to access Facebook (also removing Facebook buttons from any website), ii) redirect all requests from Google search to Bing search, iii) block the usage of adult sites, Ebay, Amazon, and YouTube, and iv) all HTTP(S) requests exchanged with Dropbox and Twitter are reported to the corporate \collector.

\section{Preliminary Prototype}
\label{sec:exres}

We develop a preliminary \X prototype in which we implement the \X layer as a Firefox plugin. It supports rules, and the \texttt{block}, \texttt{redirect}, \texttt{log\&report} actions.\footnote{For instance, the configurations developed for the Corporate and Kid profiles are available at \url{https://db.tt/yIl0LyX1}.} The \collector is a Java-based web service, which communicates with clients using SOAP. During the registration phase, the \collector provides the \X instance with a randomly assigned ID. The \collector component receives and stores reports generated by the various \X plugins.  The \X plugin has been installed successfully on both the desktop and the Android version of Firefox. In the following we present simple experiments collected using this prototype. 

\begin{figure}[t!]
    \centering
    \includegraphics[width=0.9\columnwidth]{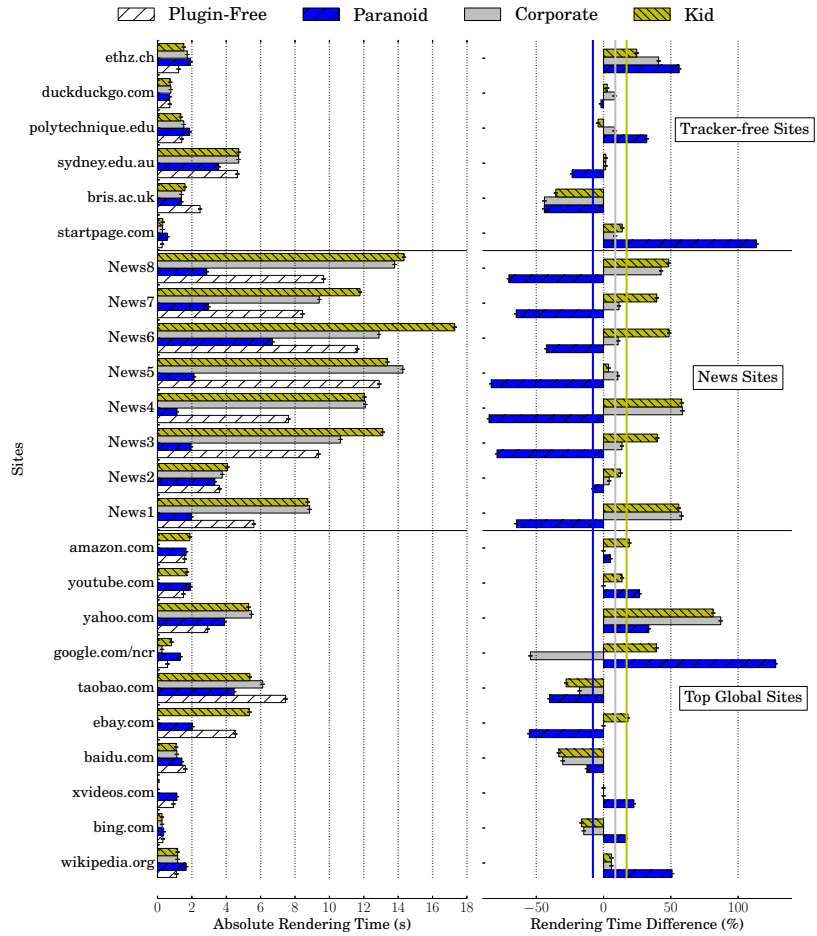}
    \vspace*{-0.5cm}
    \caption{Page rendering time cost for different plugin setups. Absolute numbers in left-hand plot, relative values with respect to the Plugin-free setup in the right-hand plot.}
    \label{fig:total_loading_time}
    \vspace*{-0.7cm}
\end{figure}

\vspace*{-0.3cm}
\subsection{Processing Overhead}
\label{sec:overhead}
First, we evaluate the performance overhead a user would pay when running the \X plugin. Given the not-optimized implementation of the prototype, these benchmarks are meant to show the feasibility of the approach rather than being considered as a thorough testing. 
We consider the three profiles described in Sec.~\ref{sec:examples}. For baseline, we take a plugin-free configuration.
We setup a testbed based on Selenium WebDriver\footnote{\url{http://www.seleniumhq.org}} to automatize the browsing of a selected set of webpages. In particular, we consider i) the Alexa top 10 global websites, ii) 8 news portals, and iii) 6 portals which do not include any online tracker.
We run the experiment from a standard PC and instrument the browser to visit each website 20 times. After discarding the best and the worst samples, we measure the average time needed to render the webpage. We purge the browser cache and cookies after each visit.

The left-hand plot in Fig.~\ref{fig:total_loading_time} reports the average rendering time for each website and for each profile. We observe that news portals are the slowest at rendering, most of them taking more than 8~s to fetch and render all the content they embed. Differently, other websites show a much simpler design and their content (mostly being HTML, CSS and Javascript files) are very fast to download.
Right plot of Fig.~\ref{fig:total_loading_time} reports the relative average rendering time of each profile with with respect to the Plugin-free configuration.
For some webpages as \textit{startpage.com} and \textit{wikipedia.org}, the rendering time is very short (order of tens of ms). Thus, the relative difference among the three profiles is broadened, but it is very small in absolute numbers.  
For the case of \textit{google.com}, the Corporate profile shows much better performance than the Paranoid, since in the former profile requests are redirected to \textit{bing.com}, which in our measurements is faster at rendering.
In general, the Paranoid profile is favored, as it blocks advertisement and some Javascript content download, thus speeding up the rendering of the webpage in many cases.
The horizontal lines show the average of the relative rendering time for the three profiles. The Paranoid is 1.07 times faster than the baseline. Corporate and Kid configurations show slightly worse performance being 1.08 and 1.17 times slower, respectively. 

In summary, results are variable, with more complicated pages that suffer some extra computational costs incurred by \X plugin that has to consider and check all links. Nonetheless, being the current implementation not optimized, results hint that clients today have enough power to easily handle the extra load generated by a possible \X implementation.


\vspace*{-0.2cm}
\section{Motivations for Having \X}
\label{sec:related}
\begin{figure}[t!]
	\centering
	\includegraphics[width=0.85\columnwidth]{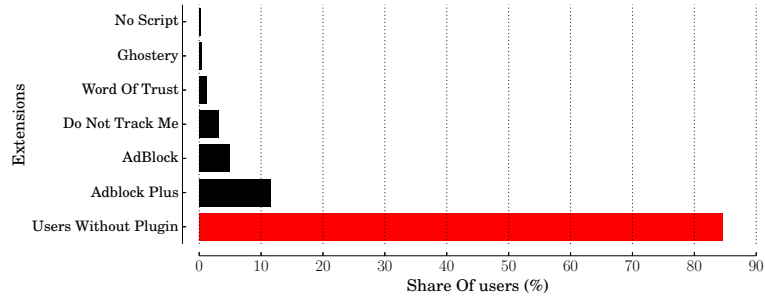}
	\vspace*{-0.4cm}
	\caption{Shares of users adopting ``popular'' privacy-preserving extensions.}
	\label{fig:extensions}
	\vspace*{-0.4cm}
\end{figure}


To demonstrate the need of a system like \X, we present some measurement facts. We analyze a 10-day long traffic trace we obtain from a large European ISP collected during October 2014 using Tstat~\cite{tstat}. To analyze both HTTP and HTTPS communications, the dataset includes anonymized TCP logs from more than 19,000 households identified by the modem IP addresses, out of which 11,000 are active (as those IP addresses from which we see at least one HTTPS request and 1000 TCP flows in the trace). We leverage DN-Hunter, a technique that allows to annotate TCP flows with original server hostname~\cite{dn-hunter}

\begin{figure}[t!]
	\centering
	\includegraphics[width=0.9\columnwidth]{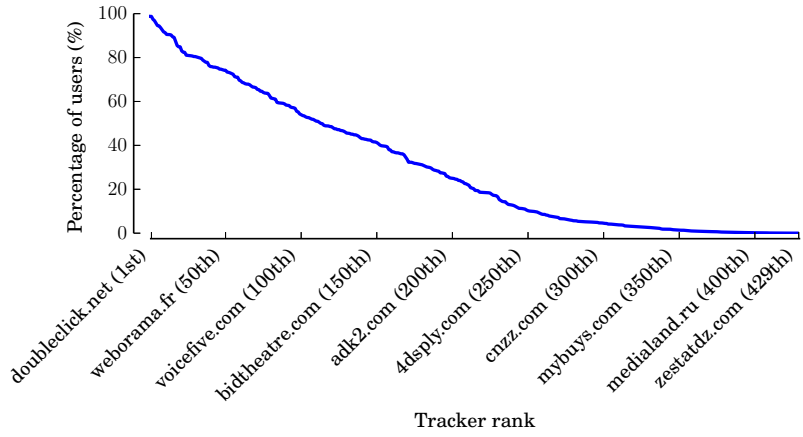}
	\vspace*{-0.2cm}
	\caption{Percentage of users contacted from top third party tracking services.}
	\label{fig:trackersDiffusion}
	\vspace*{-0.4cm}
\end{figure}

\vspace*{-0.2cm}
\subsection{Pervasiveness of Tracking Services}

We first observe how many of those users are running any plugin that could help customizing their web browsing privacy. To measure this, for each plugin, we run some active experiments to look for connections toward some update hostname. We then count the fraction of users that contacted such service for updates, and report the results in Fig.~\ref{fig:extensions}. The numbers are puzzling: only 3.1\% of users have installed DoNotTrackMe, 5\% use AdBlock, and 11.5\% use AdBlockPlus. Moreover, more than 84.5\% of users do not run any extension to limit advertisement or prevent connections to online trackers. On the other hand, we measure the pervasiveness of most popular online trackers. We build a list of more than 440 tracking services using Ghostery database. We look for users that contact them, i.e., which establish TCP connections toward tracker hostnames. The results illustrated in Fig.~\ref{fig:trackersDiffusion} are impressive: 98.8\%, 98.7\% and 97.4\% of users regularly (and unintentionally) contact top third party tracking services, i.e., DoubleClick, Google Analytics, and Google Syndication. We count 120 third party services that are contacted by more than 50\% of population. Similarly, 96.6\% and 92.4\% of users contact {Facebook} and {Twitter} (and the tracking services they include) on a daily basis, often involuntarily via the social network buttons embedded on other web pages.

These facts clearly testify how pervasive are tracking services in today Internet, and how users are unaware of their presence.

\vspace*{-0.3cm}
\subsection{Checking HTTPS Information Handling}

We run a second experiment to verify which data is sent over HTTPS when entering personal information such as user credentials and credit card data on legitimate websites.
We collect a dataset by browsing a catalog of websites with a \X enabled browser which logs all HTTP and HTTPS requests it observes. We consider a list made of the Alexa top site in the Global,  Banking, Gambling and Shopping categories. We investigate a total of 160 top sites.
For each website, we manually attempt to log in with the dummy credentials ``MyName:MyPassword''. 
Then, from the collected logs, we check how the client sends those credentials to the servers.

We find that still 10\% among the most popular websites in the global rank do not use HTTPS to exchange users' credentials.
Only 2 of these apply some custom encryption/obfuscation technique before transmitting them to the server.
Even more surprisingly, among the websites embracing HTTPS in the Global category, we notice that users' credentials are always sent in plain text over the encrypted channel. Assuming HTTPS offers a secure channel, no guarantees are given on how the server handles and stores credentials.
Indeed, the server could store those in plain text, posing severe security risks if the server gets compromised. Unfortunately, this is not a rare event. The most recent incident involved a giant like eBay~\cite{ebay}.
Even for the Bank category, 75\% of websites transmit credentials in plain text, totally trusting the HTTPS channel. Interesting, some of those do implement two-step strong authentication methods based on pins or tokens, sent in plain text through the HTTPS channel.
Similarly, 90\% of both Gambling and Shopping categories do not hash the credentials. 


These findings strengthen the need for \X to warn users about the weaknesses that are unfortunately present on (popular) websites they are used to log in. 
\vspace*{-0.3cm}




\begin{algorithm}[t!]
\scriptsize
\begin{algorithmic}[1]
\Statex \hspace{-.4cm}\textbf{Input:} $\mathbf{HS}$, $W$ \hfill \# HTTP request log and target website
\Statex \hspace{-.4cm}\textbf{Output:} $\mathbf{TS}$ \hfill \# List of possible third party trackers and their user-tracking keys
\State $\mathbf{H_{ipa}} \gets$ init\_hash\_table() \hfill \# Init hashtable of IP addresses
\State $\mathbf{H_{k-v}} \gets$ init\_hash\_table() \hfill \# Init hashtable of key-value pairs

\While{h in $\mathbf{HS}$} \hfill \# Read HTTP request logs
	\State $h \gets$ ipa, hostname, path, referer \hfill \# Extract fields of interest
	\If{$W$ not in {h.hostname} and $W$ in h.referer} \hfill \# Check target is third party 
		\State $K$, $V \gets$ extract\_keys(h.path) \hfill \# Extract keys and values from the path field
		\While{k, v in $K$, $V$} \hfill \# Iterate all key names and values
			\State $hostname\_key\_ipa \gets$ create\_hash(h.hostname, k, h.ipa) \hfill \# Create hash for $\mathbf{H_{ipa}}$
			\State $hostname\_key\_value \gets$ create\_hash(h.hostname, k, v) \hfill \# Create hash for $\mathbf{H_{k-v}}$
			\State \Call{add\_distinct}{$\mathbf{H_{ipa}}$[$hostname\_key\_ipa$], v} \hfill \# Insert all key-value pairs in $\mathbf{H_{ipa}}$
			\State \Call{add\_distinct}{$\mathbf{H_{k-v}}$[$hostname\_key\_value$], h.ipa} \hfill \# Insert the IP address in $\mathbf{H_{k-v}}$
		\EndWhile
	\EndIf
\EndWhile
\Statex 
\While{hash in $\mathbf{H_{ipa}}$} \hfill \# Iterate over $\mathbf{H_{ipa}}$ 
	\While{value in $\mathbf{H_{ipa}}$[hash]} \hfill \# Iterate over values mapped to current hash
		\If{\Call{len}{$\mathbf{H_{ipa}}$[hash]} == 1} \hfill \# Check current hash refers to one value only
			\State $hostname$, $key$, $ipa \gets$ decode\_hash(hash) \# Decode hash into hostname, key and IP address
			\State $hash_{aux} \gets$ create\_hash($hostname$, $key$, $value$) \hfill \# Create an auxiliary hash using hostname, key and value
			\If{\Call{len}{$\mathbf{H_{k-v}}$[$hash_{aux}$]} == 1 and $\mathbf{H_{k-v}}$[$hash_{aux}$] == $ipa$} \hfill \# Check the auxiliary hash in $\mathbf{H_{k-v}}$ contains only one IP address and check this corresponds to the one in $\mathbf{H_{ipa}}$
				\State \Call{add}{$\mathbf{TS}$, $hostname$, $key$} \hfill \# Add hostname and key to the output list
			\EndIf
		\EndIf
	\EndWhile
\EndWhile
\end{algorithmic}
\caption{Automatic third party tracker identifier.}
\label{alg:aut_algorithm}
\end{algorithm}

\section{Automatic Detection of Tracker: a Simple Algorithm}
\label{sec:algorithms}

One of the design challenges of \X is the need of automatic means to detect services that possibly offend users' privacy. This section presents a simple preliminary solution to automate advice generation. Specifically, we present an unsupervised methodology for an automatic \controller to identify possible third party trackers that users unknowingly contact while browsing a given website. 


We consider the set of HTTP requests that a user generates when visiting a {\it target} website. The \X layer running in user's device monitors the HTTP traffic and sends to the \collector the anonymized user identifier, and a sample of HTTP request logs having i) the hostname different from {\it target}, and ii) {\it target} appearing in the referer field. In other words, \X reports to the \collector all the third party URLs a user contacts when accessing the webpage {\it target}. Given this input, the \controller looks for parameters in the URLs that may suggest the third party service is using some identifier to track the users.

Our algorithm, illustrated in Alg.~\ref{alg:aut_algorithm}, extracts all HTTP parameters  from the third party URLs. For example, from the third party URL {\tt http://www.acme.com/query?key1=X\&key2=Y}, it extracts {\tt key1} and {\tt key2}, with values {\tt X} and {\tt Y}, respectively; {\tt www.acme.com} is the third party hostname. For each hostname and for each key, we investigate one-to-one mapping between the \X user identifier and the observed values.
Intuitively, we look for keys whose value is uniquely associated to the user.
This hints to the key being a ``user identifier'', and thus the algorithm labels the third party hostname as ``tracker''.

\begin{table}[t!]
    \centering
    \resizebox{0.5\columnwidth}{!}{
    \begin{centering}
    \begin{tabular}{c | l | l }
    \textbf{Website}                    & \textbf{Third party hostname}          & \textbf{Keys} \\ \hline
    \multirow{3}{*}{News1}     & \textit{pix04.revsci.net} & \texttt{id} \\
                                        & \textit{su.addthis.com}   & \texttt{puid} \\
                                        & \textit{track.adform.net} & \texttt{icid} \\ \hline
    \multirow{5}{*}{YouTube}   & \textit{bh.ams.contextweb.com} & \texttt{vgd} \\
                                        & \textit{eu-jet-01.sociomantic.com}   & \texttt{fpc} \\
                                        & \textit{ib.adnxs.com} & \texttt{uuid} \\
                                        & \textit{uip.semasio.net} & \texttt{sExtCookieId} \\
                                        & \textit{www.wajam.com} & \texttt{install\_timestamp} \\  \hline
    \multirow{7}{*}{Facebook}  & \textit{adadvisor.net}    & \texttt{bk\_uuid} \\
                                        & \textit{data.bncnt.com}   & \texttt{uid} \\
                                        & \textit{go.flx1.com} & \texttt{anuid, euid} \\
                                        & \textit{ira.spysomeone.com} & \texttt{s} \\
                                        & \textit{tags.bluekai.com} & \texttt{google\_gid} \\
                                        & \textit{ww1.collserve.com} & \texttt{bk\_uuid} \\
                                        & \textit{www.skyscanner.com} & \texttt{ksh\_id} \\
    \end{tabular}
    \end{centering}}
    \vspace{0.3cm}
    \caption{The third party trackers and the user-tracking keys we find in our HTTP trace for the targets News1, YouTube and Facebook.}
    \label{tab:keys}
    \vspace*{-0.4cm}
\end{table}

We validate our approach using our passive traffic trace, which contains enough data to pinpoint trackers. We target three popular web portals, News1, YouTube, and Facebook. We check all third party hostnames, and we extract those keys whose values show a one-to-one mapping with the IP addresses of the client which is considered an user identifier in our traces.\footnote{The ISP assigns a static IP address to each household modem.} In this experiment, we consider those keys for which we observe at least 25 distinct IP addresses. 

Thus, we run the algorithm to pinpoint the possible third party trackers, and the keys they employ to store the users' identifiers. Results are reported in Table~\ref{tab:keys}. As shown, we identify 3, 5 and 8 trackers for News1, YouTube and Facebook, respectively. Keys clearly suggest the exchange of possible user identifiers. The only possible false positive is \texttt{install\_timestamp}, which however we verify manually to be a unique user identifier. Looking at the hostnames, most of them are known tracking services that already appear in our list. The only exception is \textit{www.skyscanner.com} (a flight booking website) which tracks the users using the key \texttt{ksh\_id}.

Results are promising and show that the availability of large data enables automatic detection of personal leakage information. While the our experiments are carried over HTTP traffic, \X allows us to check for privacy leakage also on HTTPS connection, by checking other key-value pair, e.g., in cookies.


\section{A Feasibility Check}
\label{sec:feasibility}

\begin{figure}[t!]
    \centering
    \includegraphics[width=0.7\columnwidth]{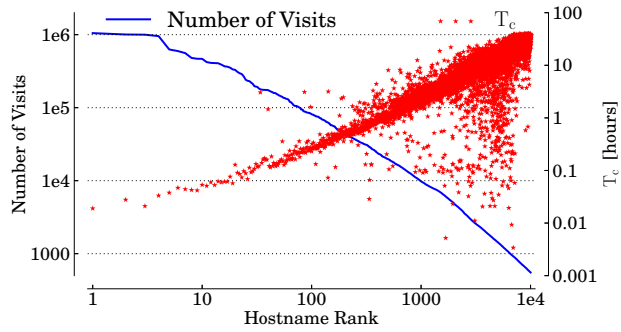}
    \caption{Popularity of hostnames found in a portion of our trace, and corresponding average time $T_c$ (in hours) to collect at least 100 with a sample ratio equal to $1/10$.}
    \label{fig:rank}
    \vspace*{-0.3cm}
\end{figure}

As described in Sec.~\ref{sec:system}, the crowd feedback is vital for \X. Therefore, we study how long the system should take to build a large enough data collection to get reliable analysis results to generate the advices. We consider for instance the case in which we aim at collecting data involving $N$ different hostnames. We say a hostname data to be reliable when we have collected at least $K$ entries, and we assume that an entry is reported to the \collector when a user visits such hostname. We apply the sampling so that only a fraction of entries are eventually seen by \X.

Understanding how many visits we need to obtain $K$ samples for each of the $N$ hostnames is a problem that belongs to Coupon Collector's family. 
In particular, we have to refer to the Newman-Shepp generalization~\cite{newman1960double}. In this case, the expectation $E[V]$ of the number of visits $V$ needed to collect a constant number of entries $K$ for a large number of hostnames $N$ is given by:
\vspace*{-0.2cm}\begin{equation}
E [V] = N \log N + (K - 1) N \log \log N + O(N).
\label{eq:coupon}
\end{equation}
The model assumes visits are equally distributed among $N$ hostnames. Considering the trace described in Sec.~\ref{sec:related} where $19,000$ households contacts every day more than $290,000$ distinct hostnames. By combining this data with Eq.~\ref{eq:coupon}, we obtain that the system would take only $8$ days to collect at least $K$=$100$ reports for each of the $N$=$10,000$ selected hostnames in the catalogue. 

In reality, the probability of visiting an hostnames typically follows a heavy-tailed Zipf-like distribution. For instance, on the left y-axis, Fig.~\ref{fig:rank} reports the number of visits of top $10,000$ hostnames. As expected, it follows the typical Zipf-like distribution. Thanks to this, the top $10,000$ hostnames correspond to 88.13\% of total visits. Therefore, we run a trace-driven experiment using the actual trace to evaluate the average time $T_c$ needed to collect $100$ samples for each of the top $10,000$ hostnames in the trace.
We assume \X clients are configured with sampling ratio equal to $1/10$. We focus on the top $10,000$ popular hostnames in the first day of our trace.
For each request, we measure $T_c$, averaging over 12 independent runs, i.e., starting the collection at a random time. As soon as  $K$=$100$ samples are collected, the hostname advice is said to be reliable.
 
The right y-axis of Fig.~\ref{fig:rank} reports $T_c$ (in hours) needed to reach the minimum critical mass of $K$=$100$ visits. As shown, \X would take few seconds to collect 100 visits for the most popular hostname (e.g., $87$~s for \textit{www.google.com}). Less that $48$~h are needed in the worst case. Observe that some services show very bursty traffic patterns that considerably decrease $T_c$. Indeed, when the clients access those services, we collect a large number of samples in few time. The overall average value of $T_c$ is $12.57$~h, much less than the time predicted by Eq.~\ref{eq:coupon}.

This simple experiment clearly shows that even with a population of only $19,000$ contributors that are reporting $1/10$ of their activity we can easily collect enough data to compute advices. We can also envision smarter sampling policies to, e.g., avoid to keep collecting samples from most popular sites while only asking sample contributions for other services.

\section{Discussion and Future Work}
\label{conclusion}

This paper presented \X, a novel crowd-sourced holistic approach to empower informed choices in the web. Motivated by the fact that today service owners have the (almost total) control on information they can collect, and by the fact that users and companies are more and more kept out of the loop, we advocate the need for any user, any device, any network to have the freedom to control the information exchanged on the Internet.

In this paper, we have shown that \X is feasible. We presented real data to show how easy would be building a crowd-sourced knowledge, supported by automatic algorithms. As a proof of concept, we implemented \X as a Firefox plugin, showing that its benefits can come at a marginal performance cost for the user. 

\X design presents some practical challenges that must be faced, and ingenuity must be used to find appropriate solutions. The research community as a whole is called to design efficient algorithms, and propose scalable implementations. \X offers this possibility, allowing anyone to contribute.

We are aware that our idea is ambitious, as, first, \X shall pass through a long and difficult standardization process to get accepted as a compelling technology by the industry, and, second, it shall undergo a deep engineering analysis to convince users about its effectiveness and usability. However, as the community is becoming more and more aware that data constitutes a vital asset in modern web, we are confident that the unified solution offered by \X represents a good starting point to protect (and possibly endorse) such asset.

\bibliographystyle{splncs_srt}

\begin{thebibliography}{10}

\bibitem{facebooksued}

\newblock Facebook sued for 15\$ billion over alleged privacy infractions, \\
  \url{http://www.cnet.com/news/facebook-sued-for-15-billion-over-alleged-privacy-infractions/}

\bibitem{GoogleSafari}

\newblock Google to pay record 22.5m fine to {FTC} over {S}afari tracking, \\
  \url{http://www.theguardian.com/technology/2012/jul/10/google-fine-iphone-ipad-privacy}

\bibitem{safebrowsing}

\newblock Google SafeBrowsing, \\
  \url{https://developers.google.com/safe-browsing/}

\bibitem{ebay}

\newblock Hackers steal vast eBay user database, including passwords, \\
  \url{http://www.bdlive.co.za/world/americas/2014/05/23/hackers-steal-vast-ebay-user-database-including-passwords}

\bibitem{Agarwal:2013:PDM:2462456.2464460}
Agarwal, Y., Hall, M.:
\newblock {P}rotect{M}y{P}rivacy: {D}etecting and {M}itigating {P}rivacy
  {L}eaks on i{OS} {D}evices {U}sing {C}rowdsourcing.
\newblock In: {ACM} Mobi{S}ys. (2013)

\bibitem{dn-hunter}
Bermudez, I.N., Mellia, M., Munafo, M.M., Keralapura, R., Nucci, A.:
\newblock {DNS} to the {R}escue: {D}iscerning {C}ontent and {S}ervices in a
  {T}angled web.
\newblock In: {ACM} {IMC}. (2012)

\bibitem{dingledine2004tor}
Dingledine, R., Mathewson, N., Syverson, P.:
\newblock Tor: {T}he second-generation onion router.
\newblock In: USENIX Security Symposium. (2004)

\bibitem{Enck:2010:TIT:1924943.1924971}
Enck, W., Gilbert, P., Chun, B.G., Cox, L.P., Jung, J., McDaniel, P., Sheth,
  A.N.:
\newblock Taint{D}roid: {A}n {I}nformation-flow {T}racking {S}ystem for
  {R}ealtime {P}rivacy {M}onitoring on {S}martphones.
\newblock In: {USENIX} {OSDI}. (2010)

\bibitem{tstat}
Finamore, A., Mellia, M., Meo, M., Munafò, M.M., Rossi, D.:
\newblock Experiences of internet traffic monitoring with tstat.
\newblock {IEEE} {N}etwork (2011)

\bibitem{kramer:pnas}
Kramer, A.D.I., Guillory, J.E., Hancock, J.T.:
\newblock {Experimental evidence of massive-scale emotional contagion through
  social networks}.
\newblock {PNAS} (2014)

\bibitem{finamore:conext14}
Naylor, D., Finamore, A., Leontiadis, I., Grunenberger, Y., Mellia, M.,
  Papagiannaki, K., Steenkiste, P.:
\newblock The {C}ost of the {\textquotedblleft}{S}{\textquotedblright} in
  {HTTPS}.
\newblock In: {ACM} {C}o{NEXT}. (2014)

\bibitem{newman1960double}
Newman, D.J.:
\newblock The double dixie cup problem.
\newblock American Mathematical Monthly (1960)

\bibitem{Popa:hotnets2010}
Popa, L., Ghodsi, A., Stoica, I.:
\newblock {HTTP} {A}s the {N}arrow {W}aist of the {F}uture {I}nternet.
\newblock In: {ACM} Hot{N}ets. (2010)

\bibitem{RiedererHotnets2011}
Riederer, C., Erramilli, V., Chaintreau, A., Krishnamurthy, B., Rodriguez, P.:
\newblock For {S}ale : {Y}our {D}ata: {B}y : {Y}ou.
\newblock In: {ACM} Hot{N}ets. (2011)

\end{thebibliography}


\end{document}